\documentclass[conference]{IEEEtran}
\IEEEoverridecommandlockouts
\usepackage{cite}
\usepackage{amsmath,amssymb,amsfonts}
\usepackage{algorithmic}
\usepackage{graphicx}
\usepackage{textcomp}
\usepackage{xcolor}
\usepackage{multirow}
\def\BibTeX{{\rm B\kern-.05em{\sc i\kern-.025em b}\kern-.08em
    T\kern-.1667em\lower.7ex\hbox{E}\kern-.125emX}}
\begin{document}

\title{RITFIS: Robust input testing framework for LLMs-based intelligent software\\
}
\author{\IEEEauthorblockN{Mingxuan Xiao\IEEEauthorrefmark{1}, Yan Xiao\IEEEauthorrefmark{2}, Hai Dong\IEEEauthorrefmark{3}, 
		Shunhui Ji\IEEEauthorrefmark{1} and Pengcheng Zhang\IEEEauthorrefmark{1}} \IEEEauthorblockA{\IEEEauthorrefmark{1}College of Computer Science and Software Engineering, Hohai University, Nanjing, China\\
    xiaomx@hhu.edu.cn, shunhuiji@hhu.edu.cn, pchzhang@hhu.edu.cn} 
	\IEEEauthorblockA{\IEEEauthorrefmark{2}School of Cyber Science and Technology, Shenzhen Campus of Sun Yat-sen University, Shenzhen, China\\
	  xiaoy367@mail.sysu.edu.cn}
	\IEEEauthorblockA{\IEEEauthorrefmark{3}School of Computing Technologies, RMIT University, Melbourne, Australia,
	  hai.dong@rmit.edu.au}}

\maketitle

\begin{abstract}
The dependence of Natural Language Processing (NLP) intelligent software on Large Language Models (LLMs) is increasingly prominent, underscoring the necessity for robustness testing. Current testing methods focus solely on the robustness of LLM-based software to prompts. Given the complexity and diversity of real-world inputs, studying the robustness of LLM-based software in handling comprehensive inputs (including prompts and examples) is crucial for a thorough understanding of its performance.

To this end, this paper introduces RITFIS, a \underline{R}obust \underline{I}nput \underline{T}esting \underline{F}ramework for LLM-based \underline{I}ntelligent \underline{S}oftware. To our knowledge, RITFIS is the first framework designed to assess the robustness of LLM-based intelligent software against natural language inputs. This framework, based on given threat models and prompts, primarily defines the testing process as a combinatorial optimization problem. Successful test cases are determined by a goal function, creating a transformation space for the original examples through perturbation means, and employing a series of search methods to filter cases that meet both the testing objectives and language constraints. RITFIS, with its modular design, offers a comprehensive method for evaluating the robustness of LLM-based intelligent software.

RITFIS adapts 17 automated testing methods, originally designed for Deep Neural Network (DNN)-based intelligent software, to the LLM-based software testing scenario. It demonstrates the effectiveness of RITFIS in evaluating LLM-based intelligent software through empirical validation. However, existing methods generally have limitations, especially when dealing with lengthy texts and structurally complex threat models. Therefore, we conducted a comprehensive analysis based on five metrics and provided insightful testing method optimization strategies, benefiting both researchers and everyday users.
\end{abstract}


\section{Introduction}
Intelligent software based on large language models (LLMs), such as ChatGPT\footnote{https://openai.com/chatgpt}, New Being\footnote{https://www.bing.com}, Chatsonic\footnote{https://writesonic.com}, and Paradot\footnote{https://www.paradot.ai}, has garnered widespread attention due to its exceptional semantic understanding and text generation capabilities. Humans can collaborate with LLMs through `prompt + example' as input~\cite{zhu2023promptbench}, thereby accomplishing many security-relevant natural language processing (NLP) downstream tasks, such as financial sentiment analysis~\cite{wu2023bloomberggpt}, public opinion monitoring~\cite{espinosa2024use}, and fraud detection~\cite{jiang2024detecting}, among others. Taking financial sentiment analysis as an example, the financial market generates a massive amount of news, reports, and social media content, making manual analysis of these texts impractical. Investors and financial analysts need to understand market sentiment to predict the performance of stocks, bonds, or other financial products. Therefore, it is necessary to collect relevant financial textual data, apply models to classify the sentiments of untagged financial texts, and use this for analysis and prediction. If influenced by adversarial texts, leading to inaccurate text classification, this can result in incorrect market sentiment judgment, misguiding investors and decision-makers into making unfavorable investment decisions, potentially leading to financial losses, reputation damage, market turbulence, and other severe consequences. Additionally, in the United States alone, the labor force for software testing is estimated to cost \$48 billion annually~\cite{davis2023nanofuzz}. For the development of stable-performing LLM-based intelligent software, and to ensure the quality of software responses, automated robustness testing of such software is crucial~\cite{ouyang2023quality}.

Current research focuses on the high sensitivity of LLMs to prompts~\cite{guo2023evaluating}. Zhu et al.~\cite{zhu2023promptbench} have assessed the robustness of LLMs to adversarial prompts by dynamically creating minor perturbations in characters, words, and sentences. Liu et al.~\cite{liu2023trustworthy} have evaluated the robustness of LLMs to typographical errors in prompts using the Justice dataset. We believe that studying the overall input robustness of LLMs (prompt + example) is more important, as real-world inputs are often complex and varied~\cite{liu2023robustness}. Evaluating the model's robustness to overall input is essential to comprehensively measure its performance, including understanding complex contexts, handling various input errors, and adapting to non-standard usages.

Therefore, this paper proposes RITFIS, a robust input testing framework for LLM-based intelligent software. To our knowledge, RITFIS is the first framework designed to evaluate the robustness of LLM-based intelligent software to natural language inputs. Inspired by the work of Morris et al.~\cite{morris2020textattack}, RITFIS consists of objective functions, perturbations, constraints, and search methods. It transfers 17 automated testing methods aimed at the deep neural network (DNN)-based intelligent software to the LLM-based intelligent software testing scenario, covering character-level, word-level, sentence-level, and combination-level perturbations. Considering the strong generalization capability of large models and the fact that many companies and individual users will not or cannot fine-tune large models for specific domains due to insufficient data~\cite{head2023large}, RITFIS focuses on zero-shot situations. We conducted robustness tests on the famous open-source large model Llama~\cite{touvron2023llama} using five of the latest or commonly used DNN-based intelligent software testing methods on three datasets that include different language styles and industry terminologies. Experimental results show that current testing methods can reveal certain robustness flaws in LLM-based intelligent software, but there are limitations in testing capabilities.

\begin{figure*}[ht]
 \centering
 \includegraphics[width=0.8\linewidth]{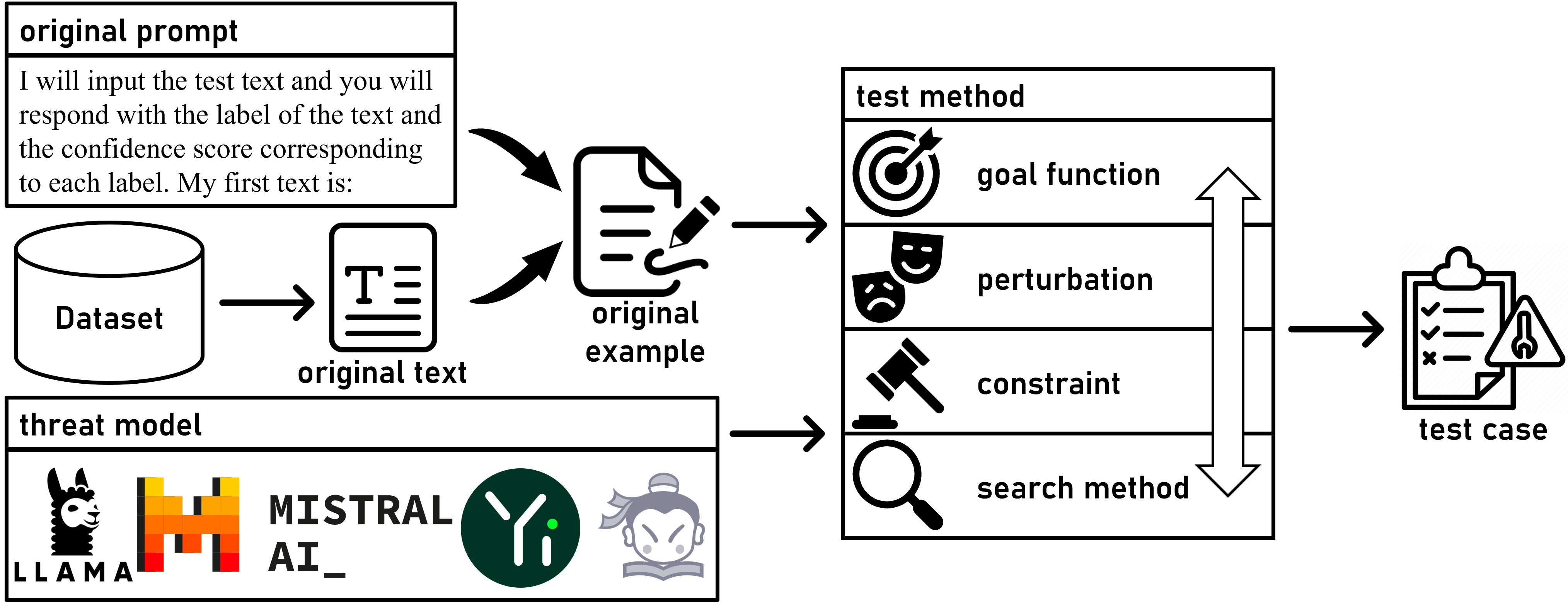}
 \caption{RITFIS workflow}\label{Fig1}
 \vspace{-0.5cm}
\end{figure*}
\section{methodology}
Figure~\ref{Fig1} outlines the proposed RITFIS, which aims to accomplish such testing: given the required threat model and prompt for testing, different testing methods are executed to automatically generate test cases for the original examples in the dataset. The testing method can be defined as a combinatorial optimization problem: using a goal function to define successful test cases, establishing a perturbation space for the original examples based on the method of perturbation, and using search methods to find a test case that meets the test objectives and certain language constraints. The modular design of RITFIS allows us to implement many different testing methods within a shared framework. The following is a detailed introduction to the basic modules of RITFIS: goal function, perturbation, constraint, and search method.
\subsection{Goal Function}
Automated testing methods for prompts are very popular in the robustness research of LLM-based intelligent software. Technically, given a dataset ${D=\left\{\left(x_{i}, y_{i}\right)\right\}_{i \in[N]}}$ and the original prompt $P$, where $x$ represents the original example in the dataset, and $y$ represents the ground truth label, the robustness test for prompts aims to mislead the threat model $f$ by perturbing each example $x$ with a given budget $C$ of $\delta$: ${\arg \max _{\delta \in C} E_{(x ; y) \in D} L[f([P+\delta, x]), y]}$. Here $L$ represents the loss function, and ${[\cdot, \cdot]}$ represents the concatenation operation. Existing work~\cite{zhu2023promptbench} suggests that LLM-based intelligent software inputs can be formulated in this way. For instance, in text classification tasks, $P$ could be ``I will input the test text and you will respond with the label of the text and the confidence score corresponding to each label. My first text is:'', and $x$ could be ``Net sales in 2007 are expected to be 10\% up on 2006.''

In this paper, our focus is on testing the overall input $[P,x]$ rather than just the prompt $P$. In the actual use of LLM-based intelligent software, both prompts and samples are indispensable, and the software often faces non-standardized or diverse inputs. Researching the robustness of the overall input helps enhance the model's robustness and security. Moreover, it is known that LLM-based intelligent software is highly sensitive to prompts~\cite{wang2023robustness,ko2023robustness}, and past robustness studies on DNN-based intelligent software~\cite{zhang2021cagfuzz,xiao2022repairing} have shown that minor perturbations to the sample can mislead the software's decisions. Therefore, research on input robustness is urgent, and we define the objective function for input robustness as follows:
\begin{equation}
\arg \max _{\delta \in C} E_{(x ; y) \in D} L[f([P, x]+\delta), y]
\end{equation}

where $\delta$ is the perturbation applied to the entirety of `prompt + example', and $L$ is the confidence associated with the label output by the threat model. This definition is similar to the generation of adversarial test cases in DNN-based intelligent software testing, and we extend this concept to the field of robustness testing for LLM-based intelligent software.
\subsection{Perturbation}
The perturbation module is responsible for directly modifying the input data of NLP models. In the robustness testing of LLM-based intelligent software, the purpose of perturbation methods is to subtly yet effectively alter the input text to probe the software's robustness. Specifically, the types of perturbations in RITFIS include: synonym replacement; confusion word replacement; insertion, deletion, and swapping of words and characters; back-translation; and template-based transformation. Taking the commonly used synonym replacement as an example, the goal of this method is to maintain the original meaning while changing the input to test the model's understanding ability, for instance, replacing `fast' with `rapid.' These transformation methods can be used individually or in combination to generate potential test cases. In RITFIS, the implementation and application of these perturbation methods are highly flexible, allowing users to customize them according to specific tasks and objectives.
\subsection{Constraint}
In RITFIS, constraints are rules or standards used to ensure that the generated test cases remain consistent with the original examples in specific aspects, and these constraints are crucial for ensuring the effectiveness and practicality of the test cases. Common constraints include: stop-word filtering, part-of-speech constraints, maximum change rate constraints, blacklist vocabulary constraints, perturbation number limits, etc. These constraints in RITFIS are configurable, allowing researchers and testers to adjust them according to specific application scenarios and requirements. Properly applying these constraints can ensure that the generated test cases can effectively test the robustness of intelligent software while maintaining acceptability to human users.
\subsection{Search Method}
Search methods are responsible for strategically exploring the potential test case space formed by perturbations, to identify successful test cases that can maximize the error in the target model. The core of search methods lies in their optimization ability, that is, intelligently applying perturbations and assessing their impact on the model output, to locate effective perturbation positions. These methods must effectively cross the model's decision boundaries while maintaining the practicality and reasonableness of the input data. The choice and implementation of search methods depend on various factors, including the specific nature of the threat model, the application scenario of the testing work, and the available computational resources. The effectiveness of search methods directly impacts the success rate and efficiency of testing, and is a key component in assessing the robustness of intelligent software~\cite{xiao2023leap}. Depending on the tester's understanding of the threat model, search methods can be classified and introduced from two perspectives: 

\begin{itemize}
\item White-box search methods require access to the model's internal structure and parameters, utilizing this information to guide the generation of test cases. These methods are usually more efficient but require an in-depth understanding of the model. Both types of search methods have their advantages and limitations.

\item Black-box search methods do not require access to the internal structure or parameters of the threat model during the testing process and rely solely on the output of the intelligent software. Black-box methods are closer to real-world application scenarios, as testers often cannot obtain detailed information about the threat model.
\end{itemize}

In RITFIS, these methods provide a diverse range of options for testing the robustness of intelligent software. White-box methods can precisely locate effective test cases when the internal information of the model is accessible, but their applicability is limited by the level of understanding of the intelligent software. In practical applications, intelligent software typically interacts with LLMs through API interfaces, without access to their internal structure and parameters. In such cases, the interaction between intelligent software and LLMs is essentially a black-box environment. RITFIS focuses on black-box search methods, allowing intelligent software to operate based solely on model output, without needing to understand the specific workings or internal mechanisms of LLMs. Focusing on black-box search methods is not only a practical choice but also a key strategy to ensure the efficiency and robustness of intelligent software in a wide range of application environments, providing a way to effectively explore the capabilities of LLMs without in-depth knowledge of the model.

\section{Experimental setup}
To assess the testing performance of existing automated testing methods on LLM-based intelligent software, we implemented 17 robustness testing methods through RITFIS. We initially evaluated the performance and efficiency of all methods, then selected five commonly used or latest testing methods for further experiments and analysis: TextFooler~\cite{jin2020bert}, StressTest~\cite{naik2018stress}, Checklist~\cite{ribeiro2020beyond}, TextBugger~\cite{li2019textbugger}, and PWWS~\cite{ren2019generating}. Experiments were conducted on three datasets, covering three scenarios: financial sentiment analysis (Financial)~\cite{malo2014good}, movie review analysis (MR)\footnote{https://huggingface.co/datasets/MrbBakh/Rotten\_Tomatoes}, and news classification (AG's News)~\cite{zhang2015character}, encompassing binary and multiple classifications as well as different text lengths. We used Llama-2-70b\footnote{https://github.com/facebookresearch/llama} as the threat model for the experiments. Composed of 70 billion parameters, Llama-2-70b demonstrates exceptional accuracy and adaptability when processing large volumes of complex data, making it an important milestone in the current field of NLP. The evaluation metrics of the experiment include:

1) \emph{Success rate} (\emph{S-rate})~\cite{morris2020textattack}, which represents the proportion of usable test cases generated by the test method among all tested examples. In this experiment, its formula can be expressed as follows:
\begin{equation}
\text{S-rate}=\frac{N_{suc}}{N}
\end{equation}
where, $N_{suc}$ is the number of test cases that test threat models successfully, and $N$ is the total number of input examples ($N$ = 1,000 in our experiment) for the current test method.

2) \emph{Change rate} (\emph{C-rate})~\cite{morris2020textattack}, which represents the average proportion of the changed words in the original text. C-rate can be expressed as:
\begin{equation}
\text {C-rate }=\frac{1}{N_{suc}} \sum_{k=1}^{N_{suc}} \frac{\operatorname{diff} T_k}{\operatorname{len}\left(T_k\right)}
\end{equation}
where $\operatorname{diff} T_k$ represents the number of words replaced in the input text $T_k$ and $\operatorname{len}$($*$) represents the sequence length. C-rate is an indicator designed to measure the difference in content between the generated test cases and the original examples.

3) \emph{Perplexity} (\emph{PPL})~\cite{morris2020textattack}, an indicator used to assess the fluency of textual test cases. Perplexity is defined as the exponentiated average negative log-likelihood of a sequence. If we have a tokenized sequence $X$=($x_0$,$x_1$,\dots,$x_t$), then the perplexity of $X$ is,
\begin{equation}
\operatorname{PPL}(X)=\exp \left\{-\frac{1}{t} \sum_i^t \log p_\theta\left(x_i \mid x_{<i}\right)\right\}
\end{equation}
where $\log p_\theta\left(x_i \mid x_{<i}\right)$ is the log-likelihood of the $i$-th token conditioned on the preceding tokens $x_{<i}$ according to the language model. Intuitively, given the language model for computing PPL, the more fluent the test case, the less confusing it is.

4) \emph{Time overhead} (\emph{T-O})~\cite{morris2020textattack}, which refers to the average time it takes for a test method to generate a successful test case.

5) \emph{Query number} (\emph{Q-N})~\cite{morris2020textattack}, which indicates the average number of times a population-based method is needed to query the threat model when generating a test case. The query number and the time overhead together reflect the efficiency of the test method.

\section{Results and analysis}
In this study, we chose the text classification task as the focus for testing the input robustness of LLMs. Text classification, as a widely applied downstream task in the NLP field, aims to assign textual content to predefined category labels, such as `positive' or `negative'. Recently observed, with LLMs outperforming traditional DNNs~\cite{sun2023text,li2023large} and crowdsourced workers~\cite{gilardi2023chatgpt,he2023annollm} in text classification, an increasing number of enterprises are inclined to use LLM-based intelligent software for text classification tasks. Table~\ref{tab1} presents the test results of five robustness testing methods on three datasets for 1000 randomly extracted original examples, with the best value for each metric highlighted in bold.
\begin{table}[h]
\centering
\caption{Performance of five methods to generate test cases.
}
\setlength\tabcolsep{4pt}
\label{tab1}
\begin{tabular}{c|c|ccccc}
\hline
\multirow{2}{*}{Dataset}   & \multirow{2}{*}{Method} & \multicolumn{5}{c}{Indicator}                                                         \\ \cline{3-7} 
                           &                         & S-rate          & C-rate         & PPL             & T-O             & Q-N            \\ \hline
\multirow{5}{*}{Financial} & TextFooler              & 0.652          & 0.012          & 51.943          & 339.459         & 95.372         \\
                           & StressTest              & 0.399          & 0.086          & \textbf{40.496} & \textbf{9.910}  & \textbf{2.981} \\
                           & Checklist               & 0.402          & 0.019          & 52.575          & 113.312         & 33.107         \\
                           & TextBugger              & 0.639          & 0.059          & 52.687          & 422.331         & 126.641        \\
                           & PWWS                    & \textbf{0.659} & \textbf{0.011} & 51.368          & 1419.016        & 452.283        \\ \hline
\multirow{5}{*}{AG's News} & TextFooler              & 0.481          & 0.017          & \textbf{48.587} & 1777.568        & 475.522        \\
                           & StressTest              & 0.042           & 0.038          & 49.588          & \textbf{10.871} & \textbf{2.749} \\
                           & Checklist               & 0.081           & \textbf{0.014} & 50.833          & 123.865         & 30.456         \\
                           & TextBugger              & \textbf{0.539} & 0.105          & 53.409          & 1097.442        & 306.723        \\
                           & PWWS                    & 0.519          & 0.015          & 49.739          & 2625.879        & 691.261        \\ \hline
\multirow{5}{*}{MR}        & TextFooler              & 0.638          & \textbf{0.018} & 62.861          & 1101.723        & 314.526        \\
                           & StressTest              & 0.119          & 0.082          & \textbf{47.406} & \textbf{9.908}  & \textbf{3.342} \\
                           & Checklist               & 0.238          & 0.020          & 62.479          & 123.521         & 42.481         \\
                           & TextBugger              & 0.659          & 0.019          & 61.065          & 786.444         & 307.560        \\
                           & PWWS                    & \textbf{0.741} & 0.013          & 60.181          & 1470.677        & 531.485        \\ \hline
\end{tabular}
\end{table}

Analyzing from the perspective of the effectiveness of the testing methods in uncovering robustness flaws, these methods face the same difficulty as when testing DNNs-based software: the longer the average length of the text in the dataset, the lower the success rate of testing. For example, TextFooler, a commonly used testing method, achieved a 63.856\% success rate on the shortest MR dataset, but only 48.143\% on the longest AG’s News dataset. The PWWS method averaged a 63.972\% success rate across the three datasets, performing better than other baseline testing methods. Notably, PWWS achieved an 80.308\% success rate on traditional DNNs, indicating that existing methods can reveal the robustness flaws of LLM-based software to a certain extent, but their testing capability for such software is still limited. We believe this issue primarily stems from two factors: First, the complexity and high integration of LLM-based software may make it difficult for standard robustness testing methods to fully reveal its potential flaws. Second, most existing testing methods rely on fixed perturbation paradigms, and the dynamic, continuous learning characteristics of LLM-based software may mean that test results at a given time point do not reflect its long-term robustness. To address these issues, we suggest improving the establishment of perturbation spaces and search methods in testing algorithms, tailored to LLMs' unique characteristics and behavioral patterns, thereby increasing the coverage and depth of testing. Secondly, considering the dynamic nature of LLM software, we need to adopt continuous, iterative testing methods and adaptive testing strategies to capture the behavioral changes of the software in long-term operation.

To further analyze the performance of the testing methods, this paper evaluates the quality of test cases generated by different testing methods through two indicators: change rate and text perplexity. All five testing methods misled LLMs with no more than an 11\% change rate, indicating that slight perturbations to the original samples could lead to erroneous outputs by LLMs. In terms of PPL, test cases generated by StressTest performed best, with an average PPL score of 45.830, meaning these test texts were more natural and fluent.

Besides the quality of test cases, the efficiency of testing methods is also a key focus of this study, including time cost and the number of queries. In these two metrics, StressTest performed best: it saved at least 103.402 seconds and reduced 27.707 queries to the threat model per successful test case compared to the next best method. However, this significant improvement in efficiency seems to come at the expense of the effectiveness of robustness testing, as StressTest had the lowest success rate across the three datasets. We believe this phenomenon is due to an overemphasis on rapid completion of testing, leading to a neglect of the comprehensiveness of test cases, thereby affecting the accuracy and reliability of test results. Secondly, excessively reducing threat model queries means insufficient exploration of software robustness flaws, lowering the possibility of uncovering complex issues. Therefore, when conducting robustness testing on LLM-based software, there is a need to rebalance the relationship between efficiency and effectiveness, ensuring that increasing the speed of testing does not sacrifice the depth and comprehensiveness of test cases. Specific actions can include increasing the depth of threat model queries to more comprehensively assess software robustness. Secondly, consider introducing more complex test case generation algorithms, which can enhance the coverage and quality of test cases while maintaining efficiency. We suggest adopting dynamic adjustment strategies, dynamically adjusting the allocation of time and query resources based on feedback during the testing process, thus optimizing the balance between efficiency and effectiveness.
\section{conclusion}
In this paper, we present RITFIS, which evaluates the input robustness of LLM-based intelligent software in NLP. RITFIS formalizes robustness testing as a combinatorial optimization task, consisting of goal function, perturbation, constraint, and search method, and applies existing testing methods for DNN-based intelligent software to the robustness testing of LLM-based intelligent software. We selected five testing methods and conducted experiments on three datasets across five testing metrics. The research findings indicate that current LLM-based intelligent software is not robust enough, and the flaw detection methods of existing testing approaches are also not sufficiently effective. Therefore, in future work, we plan to adopt continuous iterative testing methods and rebalance efficiency and effectiveness, targeting the unique input paradigms and hallucination issues of LLMs, to design more effective robustness testing methods.
\bibliographystyle{ieeetr}
\bibliography{ref}



\end{document}